\documentclass[%
 reprint,
 floatfix,
 amsmath,amssymb,
 aps,
]{revtex4-1}

\usepackage{graphicx}
\usepackage{dcolumn}
\usepackage{bm}
\usepackage{color}
\usepackage{lineno}
\usepackage{hyperref}

\usepackage{acronym}
\usepackage{eurosym}
\usepackage{bbding}  
\usepackage{rotating} 
\usepackage{array}
\usepackage{adjustbox}
\usepackage{xspace} 
\usepackage[table]{xcolor} 
\usepackage{makecell}

 \newcolumntype{R}[2]{%
     >{\adjustbox{angle=#1,lap=\width-(#2)}\bgroup}%
     l%
     <{\egroup}%
 }
\newcommand*\rot{\multicolumn{1}{R{60}{1em}}}
\newcolumntype{C}[1]{>{\centering\let\newline\\\arraybackslash\hspace{0pt}}m{#1}}

\newif\ifEdNotes \EdNotestrue

\newcommand\Tstrut{\rule{0pt}{2.9ex}}         
\newcommand{\epem}{\mbox{${\rm e}^+{\rm e}^-$}}

\def\etal{{\it et al.}}

\begin{document}

\preprint{LCCPEB---}

\title{The International Linear Collider \\ A European Perspective}

\author{\textbf{Prepared by}:
Philip Bambade$^1$,  Ties Behnke$^2$, Mikael Berggren$^2$, Ivanka Bozovic-Jelisavcic$^3$, Philip Burrows$^4$, Massimo Caccia$^{5}$, Paul Colas$^{6}$, Gerald Eigen$^{7}$, Lyn Evans$^{8}$, Angeles Faus-Golfe$^{1}$, Brian Foster$^{2,4}$, Juan Fuster$^{9}$, Frank Gaede$^{2}$, Christophe Grojean$^{2}$, Marek Idzik$^{10}$, Andrea Jeremie$^{11}$, Tadeusz Lesiak$^{12}$, Aharon Levy$^{13}$, Benno List$^{2}$, Jenny List$^{2}$, Joachim Mnich$^{2}$, Olivier Napoly$^{6}$, Carlo Pagani$^{14}$, Roman Poeschl$^{1}$, Francois Richard$^{1}$, Aidan Robson$^{15}$, Thomas Schoerner-Sadenius$^{2}$, Marcel Stanitzki$^2$, Steinar Stapnes$^{8}$, Maksym Titov$^{6}$, Marcel Vos$^{9}$, Nicholas Walker$^{2}$, Hans Weise$^{2}$, Marc Winter$^{16}$. }
\affiliation{\vspace{.2 cm}$^1$LAL-Orsay/CNRS, $^2$DESY, $^3$INN VINCA, Belgrade, $^4$Oxford U., $^{5}$U. Insubria, $^{6}$CEA/Irfu, U. Paris-Saclay, $^{7}$U. Bergen, $^{8}$CERN, $^{9}$IFIC, U. Valencia-CSIC, $^{10}$AGH, Krak\'ow, $^{11}$LAPP/CNRS, $^{12}$IFJPAN, Krak\'ow, $^{13}$Tel Aviv U., $^{14}$INFN, $^{15}$U. Glasgow, $^{16}$IPHC/CNRS. }




\date{\today}

\begin{abstract}

\centerline{\textbf{Abstract}}

The International Linear Collider (ILC) being proposed in Japan is an electron-positron linear collider with an initial energy of 250~GeV.
The ILC accelerator is based on the technology of superconducting radio-frequency cavities.   This technology has reached a mature stage in  the European XFEL project and is now widely used.  

The ILC will start by measuring the Higgs properties, providing high-precision and model-independent determinations of its parameters. The ILC at 250~GeV will also search for direct new physics in exotic Higgs decays and in pair-production of weakly interacting particles. The use of polarised electron and positron beams opens new capabilities and scenarios that add to the physics reach. The ILC can be upgraded to higher energy, enabling precision studies of the top quark and measurement of the top Yukawa coupling and the Higgs self-coupling. 

The international - including European - interest for the project is very strong. Europe has participated in the ILC project since its early conception and plays a major role in its present development covering most of its scientific and technological aspects: physics studies, accelerator and detectors. The potential for a wide participation of European groups and laboratories is thus high, including important opportunities for European industry. 

Following decades of technical development, R\&D, and design optimisation, the project is ready for construction and the European particle physics community, technological centers and industry are prepared to participate in this challenging endeavour.

\end{abstract}

\pacs{Valid PACS appear here}
\maketitle

\onecolumngrid
\textbf{Supporting documents web page:} 
\vspace{-2.6cm}

\url{https://ilchome.web.cern.ch/content/ilc-european-strategy-document}

\textbf{Contact persons:} James Brau (jimbrau@uoregon.edu), Juan Fuster(Juan.Fuster@ific.uv.es), Steinar Stapnes (Steinar.Stapnes@cern.ch)

\pagebreak

\pagestyle{plain}
\setcounter{page}{1}

\twocolumngrid

\section{\label{sec:intro}Introduction}

The International Linear Collider (ILC) is proposed as an \epem linear collider that can start operating at  250~GeV centre-of-mass energy. The physics case, technological readiness and political opportunities of the ILC are discussed in detail in the document: \emph{The International Collider. A global project}~\cite{ILCforESS} which is presented to the European strategy update process. The international  support for the ILC is very large and Europe has strongly participated since the early stages of the project at the beginning of this century.   The collider design and detector concepts are thus the result of nearly twenty years of R\&D. The heart of the ILC accelerator, the superconducting cavities, is based on pioneering work of the TESLA Collaboration. This technology is the basis for a number of operating free-electron-laser facilities (European XFEL at DESY), under construction (LCLS-II at SLAC) or in preparation (SHINE in Shanghai).

This document complements \cite{ILCforESS} and discusses the European expertise on the ILC project for both the accelerator and the detectors. The document is based on the report \cite{ejade-report} that was produced by the E-JADE Marie Curie project \cite{ejade}. Possible European contributions to the preparation and construction phases are presented. It is based on the assumption that the ILC will be realised in Japan with strong international participation. 
It is proposed that CERN will play a leading role in the European participation in the ILC, along the lines described in the  conclusions of the 2013 Update of the European Strategy, and also in a similar fashion to that  developed for the European participation in the US neutrino program.  

This report uses the same steps and timelines as the KEK ILC action plan~\cite{kekactionplan}. The ILC \textbf{Preparation Phase}, currently foreseen for 2019/20-2022/23, needs to be initiated by a positive statement from the Japanese government about hosting the ILC, followed by a European strategy update that ranks European participation in the ILC as a high-priority item. The Preparation Phase focuses on preparation for construction and agreement on the definition of deliverables and their allocation to regions. The European groups will concentrate on preparations for their deliverables including working with and preparing European industry. Europe and European scientists, as part of an international project team, will also participate in the overall finalisation of the design, while in parallel contributing to the work of setting up the overall structure and governance of the ILC project and of the associated laboratory.
The \textbf{Construction Phase} will start after the ILC laboratory has been established as an organisation, currently foreseen from 2023/24, and intergovernmental agreements are in place. At the current stage, only the existing capabilities of the European groups relevant for this phase can be outlined in broad terms. As mentioned above, the detailed contributions will have to be defined during the preparation phase and formalised by inter-governmental agreements.  With the completion of the foreseen 8-9 year construction phase in 2031-32, the European groups will naturally be involved in the commissioning of both the accelerator and detectors.

This document first reports the European expertise and track-record relating to the accelerator in section II. Section III describes the detectors and computing. Following this, a short discussion summarises the current political situation, suggests how a
European contribution might be organised, including the contribution from European industry and finally draws some conclusions.


\section{\label{sec:acc}Accelerator}
Europe has a very strong scientific, technological and industrial basis with which to make significant contributions to the construction of virtually 
any part of the ILC accelerator. In the following three subsections, the past and current activities, the ILC Preparation Phase and the subsequent Construction Phase will be described, from the perspective of European capabilities to participate in constructing the ILC accelerator. 
A final subsection considers the organisation of the future ILC accelerator activities on the European side. 

\subsection{ILC accelerator competence in Europe~\label{sec:acc:competence}}

The key recent and ongoing activities in Europe with high relevance for ILC are the direct European participation in the ILC Global Design Effort (GDE) and the Linear Collider Collaboration (LCC),
the European XFEL (E-XFEL) project, the European Spallation Source (ESS) superconducting linac, the European 
participation in the ATF-2~\cite{Grishanov:2006kx} at KEK for linear collider studies, the E-JADE Marie 
Curie project~\cite{ejade}, and the CLIC study~\cite{Aicheler:2012bya,Linssen:2012hp}. The following paragraphs 
summarise these briefly; a more comprehensive description can be found in ~\cite{ejade-report}. 

\begin{figure}[htbp]
\includegraphics[width=0.4825\textwidth]{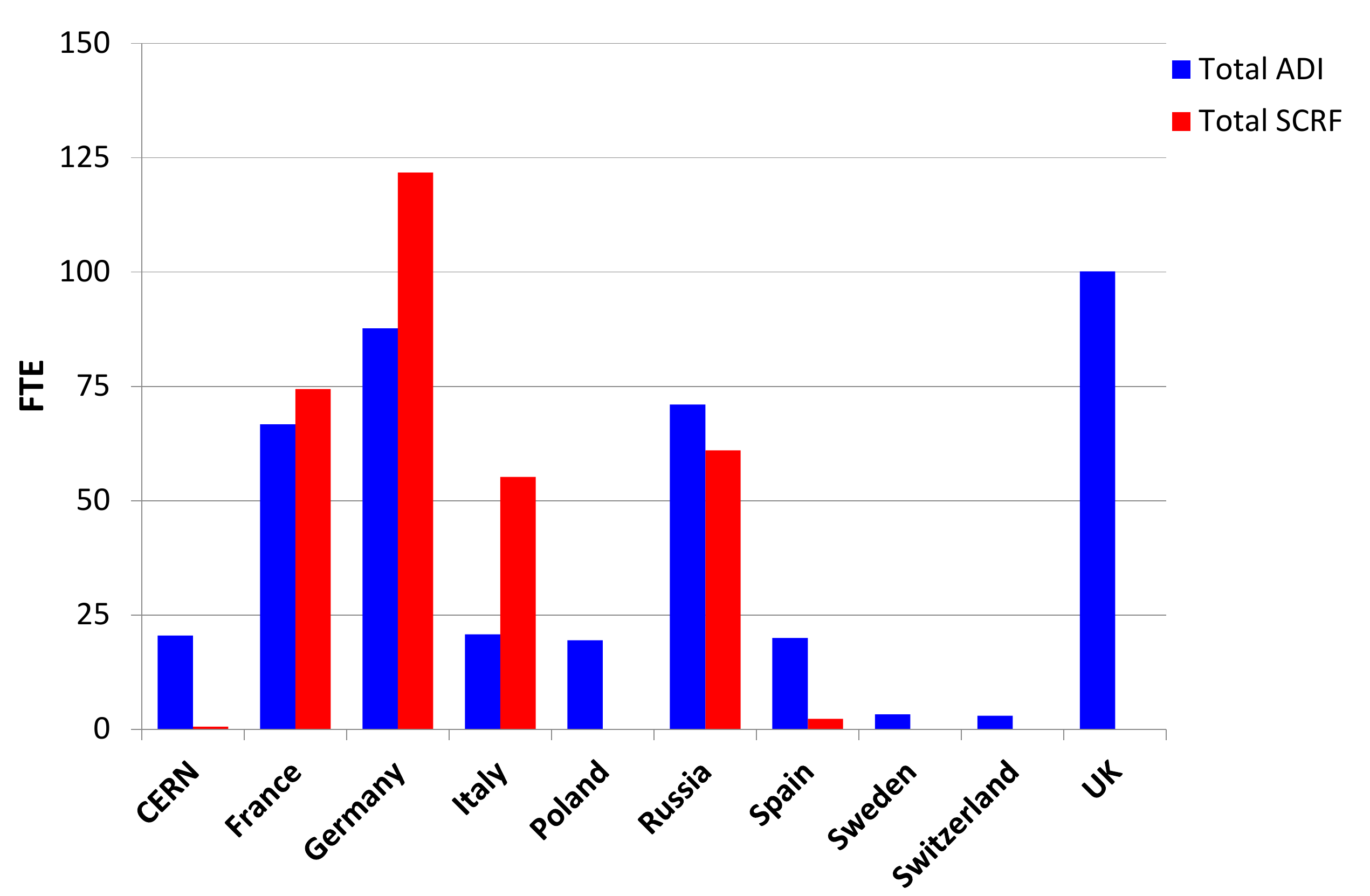}
\caption{\label{fig:PrePrep:ilcgde4} The contribution to the ILC GDE (2007-2012) in staff years per European country, separately for SCRF and Accelerator Design and Integration (ADI).}
\end{figure}

During 2007-2012, the ILC GDE was responsible for the coordination of the worldwide ILC accelerator design. 
R\&D and cost-estimate development during this period culminated in the publication of the ILC Technical Design Report in 2013~\cite{Behnke:2013xla}. 
The estimated European FTE contributions (728 person years in total) during 2007-2012 are summarised in Figure~\ref{fig:PrePrep:ilcgde4}, 
divided into accelerator domains and superconducting RF (SCRF) technology development (excluding management and documentation support). 
Note that the SCRF numbers represent ILC-specific resources and do not include the extensive synergetic contributions from the European XFEL during this period.


After the completion of the ILC TDR~\cite{Adolphsen:2013kya} in 2013, the GDE was replaced by the LCC. In spite of limited direct funding for ILC studies in Europe during the last five years, the combined efforts in several related projects have allowed European researchers to continue ILC project development activities in the framework of the LCC, in close collaboration with Japan. The most important of these projects has been the construction of the European XFEL (discussed in next paragraph). The ILC activities have been to a large degree focused on studies for implementing a 250~GeV ILC machine in Japan. These European activities are as follows:

\begin{itemize}
\item Participation in the LCC ILC-related R\&D (SCRF, Cryogenics, High-Level RF (HLRF), Civil Engineering, Beam dump, Positron source, Radiation safety)
\item Participation in the ATF-2 programme at KEK (nano-beams and final focus)
\item Combined studies with CLIC in several areas (beam-dynamics, modelling/simulation, damping rings, Damping Ring to the Main Linac transfer (RTML), Beam Delivery System(BDS), Machine-Detector Interface (MDI), cost and power)
\item E-JADE supported secondments of European researchers to Japan for ILC and ATF-2 related activities.
\end{itemize}
These activities involves groups in  France, Germany, Italy, Spain, UK and CERN. 



Beyond the activities listed above, the European XFEL project at the DESY campus is the most prominent demonstration of the European capabilities
to contribute to the construction of the ILC accelerator.
The 17.5~GeV superconducting linac at the E-XFEL comprises 97 ILC-like superconducting cryomodules, containing 776 1.3~GHz TESLA cavities
driven by 25 10-MW-multibeam klystrons. The E-XFEL linac configuration is very similar to that foreseen for the ILC and can be seen as a 7\% (in beam energy) ILC prototype. The cryomodules were produced by a consortium of six European countries together with predominantly European industries. 
 The consortium members and the various responsibilities across the linac-relevant E-XFEL work packages (including testing) are given in Table~\ref{tab:acc:XFELResponsibilities}.


\begin{table}[htbp]
\begin{tabular}{l|C{0.6cm}|C{0.6cm}|C{0.6cm}|C{0.6cm}|C{0.6cm}|C{0.6cm}|}

	& \rot{\bfseries Germany}& \rot{\bfseries France}&\rot{\bfseries Italy}&\rot{\bfseries Poland}&\rot{\bfseries Russia}&\rot{\bfseries Spain} \\\hline
\bfseries Linac&&&&&&\\[5pt]\hline
\hspace{0.3cm}Cryomodules		&\Checkmark&\Checkmark&\Checkmark&\Checkmark&&\\[5pt]
\hspace{0.3cm}SCRF Cavities		&\Checkmark&&\Checkmark&\Checkmark&&\\[5pt]
\hspace{0.3cm}Couplers and Tuners	&\Checkmark&\Checkmark&&\Checkmark&&\\[5pt]
\hspace{0.3cm}Cold Vacuum		&\Checkmark&&&&\Checkmark&\\[5pt]
\hspace{0.3cm}Cavity String Assembly	&\Checkmark&\Checkmark&&&&\\[5pt]
\hspace{0.3cm}SC Magnets		&\Checkmark&&&\Checkmark&&\Checkmark\\[5pt]\hline
\bfseries Infrastructure&&&&&&\\[5pt]\hline
\hspace{0.3cm}\makecell[l]{Accelerator Module Test\\ Facility (AMTF) }&\Checkmark&&&\Checkmark&\Checkmark&\\[5pt]
\hspace{0.3cm}Cryogenics		&\Checkmark&&&&&\\[5pt]\hline
\bfseries Sites \& Buildings&&&&&&\\[5pt]\hline
\hspace{0.3cm}AMTF hall		&\Checkmark&&&&&\\[5pt] \hline
\end{tabular}

\caption{\label{tab:acc:XFELResponsibilities} Responsibility matrix for cryomodule production and testing for the European XFEL. 
More details and a similar matrix can be found in ~\cite{ejade-report} concerning construction of SCRF modules for the ESS linac.} 
\end{table}

Construction of the European XFEL is now complete, and the SCRF linac has been brought into operation. The E-XFEL can directly benefit the ILC in several ways: 
\begin{itemize}
\item The experience and knowledge gained during the unprecedented industrial production of 100 cryomodules over a three-year period can provide 
invaluable input to any future large-scale production for the ILC. The detailed cost breakdown of the E-XFEL cryomodules 
provides a solid basis for any future projection of a possible European in-kind contribution to the ILC. 
\item 
The currently ongoing commissioning of the E-XFEL and its future operation will provide invaluable system testing for the ILC, 
including understanding the ultimate performance of the modules with beam loading, beam control (Low-level RF development), software tools, 
and more general operational experience. 
\item The infrastructure that was constructed for E-XFEL cavity and module
testing, high-power coupler conditioning and module assembly will continue to be maintained, and (in the case of the testing infrastructure) will provide a significant support for SCRF R\&D.
\end{itemize}

The expertise and infrastructures established for the E-XFEL are now partially deployed in the construction of SCRF modules for the ESS linac. 
The consortium is slightly different: Sweden and UK and several new groups from countries already involved in the E-XFEL have joined, while DESY is peripherally involved.
The ESS effort is also strategically important for ILC, as infrastructures and industry capabilities are being updated and the procedures for SCRF module production are being refined.

\subsection{ILC accelerator Preparation Phase activities in Europe ~\label{sec:acc:prephase}}

The overall resources needed during the four-year Preparation Phase are estimated to be 5\% of the material and 10\% of the personnel 
foreseen for the 250~GeV accelerator project construction~\cite{Evans:2017rvt}. Irrespective of the final level of European investment into the ILC,
it would be appropriate that, given its expertise and previous involvement, Europe invests 1/3 of the overall effort required in the 
preparation phase. The remainder of this section assumes this to be the case. This would then amount to a total European material 
budget of 85 M\euro{} and 240 FTE-years, integrated over the period~\cite{ejade-report}.
Distributed over four years, an average yearly budget of around 30 M\euro{} (covering material and personnel) would be needed with an 
increasing profile from 2019 towards 2022. The resources required for the activities in the preparation phase are hence similar in 
scope to those used by existing project studies such as CLIC and FCC. One important difference will be that the ILC preparation 
requires a strong engineering team, such that the profile of the ILC personnel will be towards more engineering and technical 
design and will gradually become less focused on R\&D studies.

The preparation phase will be used for three main purposes:

\begin{itemize}
\item 
Technical preparation of the major European deliverables foreseen for the construction phase. 
This covers final technical specifications, final prototypes, the preparation of pre-series orders and the preparation of local 
facilities. A particularly important point for Europe is the transfer of European XFEL know-how and the preparation of the 
relevant facilities for ILC construction.
\item
The second major technical activity will be the organisation of a strong European design office for ILC that will liaise with 
other such offices: there will certainly be a host-lab office in Japan, but additional international design offices will be 
required. In Europe, the installation of a central European Design Office at CERN with satellite offices in other European laboratories is considered the most viable model.
\item 
The third key activity in the preparation phase will be negotiations about the final European ILC contributions, about the 
organisation of the project in the construction and operation phase, and about a future governance model for the ILC. 
These issues are discussed in further chapters of this document.
\end{itemize}

The main technical activities are summarised in table~\ref{fig:prep-phase-summary}. 


\begin{table}[htbp]

\begin{tabular}{p{3.5cm}p{4.75cm}}
 \bfseries {Activities in Europe} &\bfseries{More details}                                                         \\[4pt]\hline\noalign{\smallskip}
SCRF activities			&Cavity fabrication and preparation, Power Couplers, Automation of assembly, E-XFEL $\rightarrow$ ILC\\[4pt]
High efficiency klystron R\&D   &Reaching 70\% efficiency or above with several vendors   \\[4pt]
Cryogenics system               &LHC system similar in size to ILC\\[4pt]
Accelerator Domain Issues       &Positron source, Damping Rings, Beam Delivery Systems, Low emittance beam transport, Beam dumps\\[4pt]
Detector and Physics            &Design optimisation, MDI, Technical prototypes, TDR, Physics Studies, Sodtware\\[4pt]
Documentation system            &Experience from E-XFEL                                       \\[4pt]
Regional Design office          &Naturally at CERN, linking to other European National Labs \\[5pt] \hline
\end{tabular}

\caption{\label{fig:prep-phase-summary} Main technical activities during the preparation phase.}
\end{table}

\subsection{ILC accelerator in-kind contributions from Europe during the ILC Construction Phase ~\label{sec:acc:constrphase}}

The actual contribution from Europe will be determined by negotiations at a 
later stage, and no formal commitments can be made at this time. The models 
chosen for in-kind contributions in this document, and discussed in this 
section, are as follows: Europe, being one of three major regions involved in 
the ILC project, delivers 1/3 of the non-CFS (civil engineering and conventional facilities) 
components of the ILC accelerator. The CFS work 
and components will naturally be constructed in and installed and 
commissioned by the host nation. The models follow recommendations given in 
the Nomura report~\cite{Nomura-eng} and are also drawn from the ILC Project Implementation 
Planning (PIP) document~\cite{ILCPIP}. The models discussed correspond to an overall 
European contribution of around 1090~M\euro{} and 1900 FTE-year to the ILC accelerator project. 

\begin{figure}[htbp]
\begin{center}
\includegraphics[width=0.4825\textwidth]{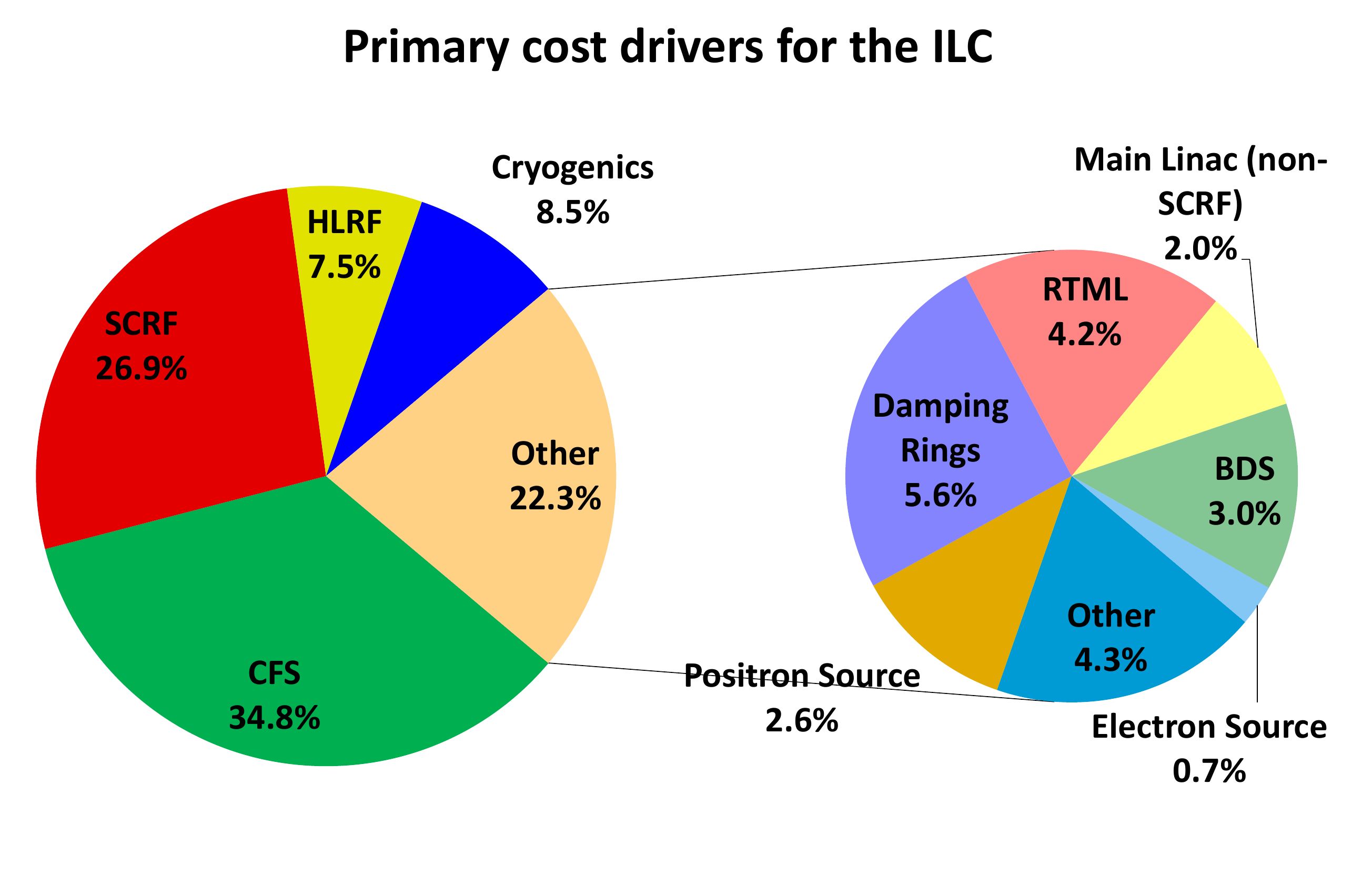}
 \caption{\label{fig:constructionmodel:ILCPrimaryCostDrivers} Primary cost drivers for the 250~GeV ILC as identified in the ILC TDR~\cite{Adolphsen:2013kya}.}
\end{center}
\end{figure}

Figure~\ref{fig:constructionmodel:ILCPrimaryCostDrivers} shows the primary cost drivers of the ILC as identified in the ILC
TDR~\cite{Adolphsen:2013kya}. The dominant cost driver is the production of approximately 930 SCRF cryomodules. Given European expertise, 
existing infrastructure, and proven industrial capability arising from the construction 
of the E-XFEL and ESS, it is widely assumed that a large fraction of the European in-kind contribution 
will be in the form of cryomodules. The ILC TDR assumed that three, possibly four, production sites 
would be required worldwide. 
The European XFEL has been constructed by a consortium of several European
countries, with DESY providing overall coordination. Based on this experience 
and the known published cost of the E-XFEL cryomodules, we have produced a model 
for producing and testing one-third of the cryomodule production (310 
cryomodules). This model has then been used to scale to other possible 
contribution scenarios discussed below. The resulting cost per cryomodule is about 
1.65~M\euro{} (material and labour), including module production and 100\% 
testing of cavities and cryomodules, which represents an approximate reduction 
of 26\% over the actual E-XFEL cost. This reduction has been estimated through 
the higher production numbers and the re-use of existing E-XFEL production and 
testing infrastructure.  Where applicable, a mild learning curve slope of 95\% 
has been applied, assuming two vendors for procurement of all major sub 
components.

Europe not only has expertise in cryomodule production, but also in many other 
of the subsystems required for the ILC. A European contribution to ILC could 
therefore include other items in addition to cryomodules. As an example, providing 
one third of the klystrons, modulators and associated controls (low-level RF) 
needed for the SCRF linacs would cost around 155~M\euro{}, one third of the cost of 
the cryogenics systems would be roughly 143~M\euro{}, and supplying a fraction of the 
accelerator components (vacuum, power supplies, magnets, computing and controls 
etc.) needed for the project could easily reach a cost of approximately 345~M\euro{}. 
Figure~\ref{fig:constructionmodel:ILCPrimaryCostDrivers} also shows the ILC TDR costs broken down by both accelerator sub-system and 
technical components, not including CFS, installation or SCRF cryomodules.

The experience of the European countries, organised according to sub-systems, including overall design studies, are shown in Table~\ref{tab:ConstructionMatrix}. There is significant potential for the countries to get involved in other activities than shown in the table, and for other European countries to get involved, based on experience with other accelerator projects than those providing the information in the table.

\begin{table*}

\begin{tabular}{l|C{1.8cm}|C{1.8cm}|C{1.8cm}|C{1.8cm}|C{1.8cm}|C{1.8cm}|C{1.8cm}|C{1.8cm}|}
  	&\bfseries SCRF	& \bfseries HLRF	&\bfseries Sources&\bfseries Damping Rings	&\bfseries Instru\-mentation&\bfseries Beam Dynamics	&\bfseries Beam Delivery System	&\bfseries Cryogenics \\\hline\Tstrut
CERN	&	&C,O	&O	&G,C,O		&C,G		&C,G		&C,G			&O\\
France	&X,E,G	&	&G	&		&A,G		&G		&C,G			&\\
Germany	&X,G	&X	&G	&G		&X		&G		&			&X,O\\
Italy	&X,E,G	&	&	&G		&		&		&			&\\
Poland	&X,E	&	&	O&		&E,O		&		&			&X,E,O\\
Russia	&X	&	&G	&		&		&		&			&\\
Spain	&X,E	&	&	&		&A		&		&C,G			&\\
Sweden	&E	&	&	&		&		&		&G			&\\
Switzerland& 	&	&	&		&X,C		&		&			&\\
UK	&E	&	&G	&G		&A,C,G		&C,G,A		&C,G,A			&\\ \hline
\end{tabular}

\caption{\label{tab:ConstructionMatrix} European expertise relevant for ILC accelerator construction, based on experience in the recent past. This is based on two major construction projects, the  E-XFEL (X) and the ESS (E), several more R\&D oriented efforts namely the GDE/LCC (G), ATF-2 (A), CLIC (C) and  experience in other accelerator projects (O)}
\end{table*}

\begin{figure}[htbp]
\includegraphics[width=0.48\textwidth]{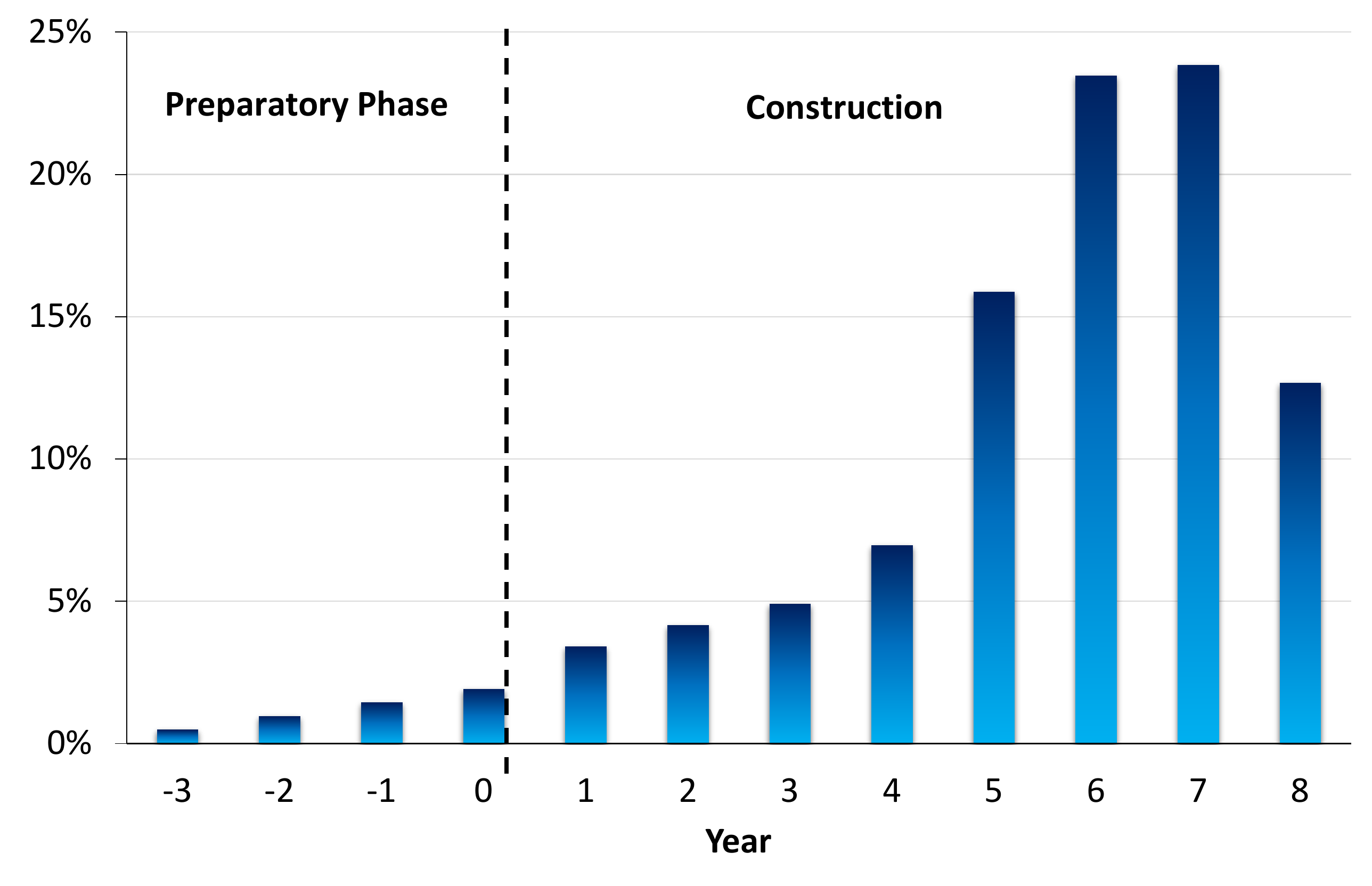}
\includegraphics[width=0.48\textwidth]{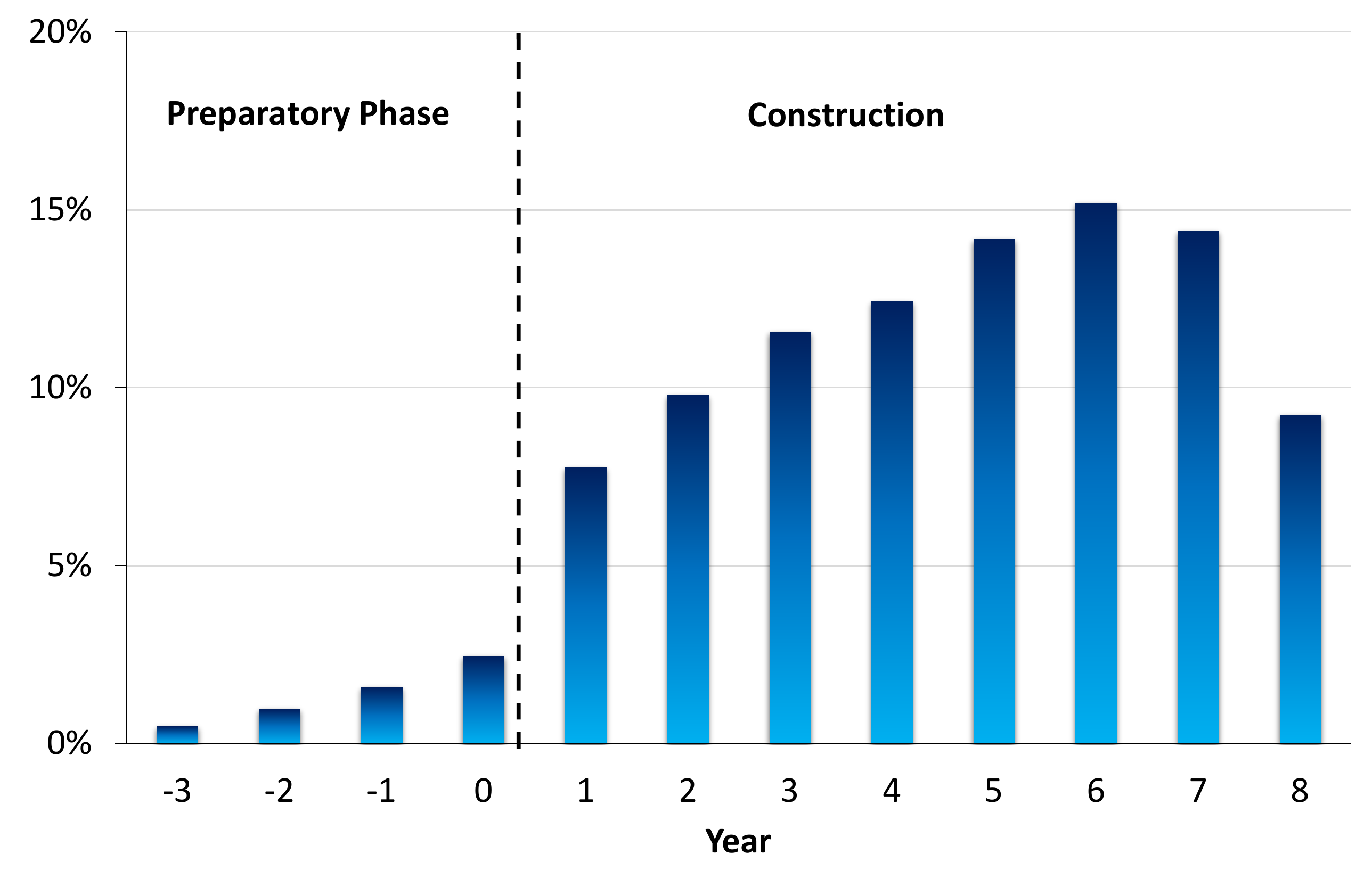}
\caption{\label{fig:costprofile:costprofile} An estimation of the cost and personnel profile cover the preparatory phase (years -3 to 0) 
and the construction phase (years 1 to 8) for the 1/3 model of a 250~GeV machine, as a fraction of the
totals. In this timeline, 
year 1 corresponds to the first year of construction, currently foreseen in 2023 
Top: material costs. Bottom: explicit personnel in FTE-years.
}
\end{figure}

As discussed in Section~\ref{sec:acc:prephase}, 5\% of the total value and 10\% of the personnel are assumed
to ramp up during the four-year preparatory phase. The profiles shown in Fig.~\ref{fig:costprofile:costprofile} also includes a fraction
of the expected lab services personnel, which will almost certainly be required during the preparatory
phase. 

\subsection{\label{sec:acc:org}Organisation of the accelerator activities}

In this short section, we discuss possible organisation forms of a European
participation in the ILC. 
Given the physics interests in a future \epem accelerator, the ILC project is likely to imply a substantial investment from
the European perspective. This fact highlights the necessity for a high-level
agreement about the level of European participation in ILC to be formalised between 2020
and 2023 if the currently assumed time-line of the ILC project is to be respected.
On the accelerator side, the typical organisational model during the construction phase is based on a leading host laboratory supported by 
bi-lateral agreements for in-kind deliverables from other funding agencies and laboratories. This was the model used during the LHC construction. 
The exact organisation of Europe's contribution to the ILC is not clear as of today and beyond the scope of this document. 
As documented above, there is a very strong potential for European contributions to the ILC accelerator construction.
To organise a broad scientific and technical European ILC effort, involving coherent contributions from many participating institutes and countries, 
seems to be the most realistic and cost-effective scenario if executed with CERN in a leading coordinating role.

\section{\label{sec:det}Detectors}

Detectors at colliding beam facilities are large international efforts, realised by strong collaborations. 
Typically many countries join forces, and contribute mostly in-kind to the building of the detector systems. 

Within the ILC project, the community has come together behind two complementary detector concepts, the SiD and the ILD concept. 
After a call for ``letters of intent'' in 2009, the proposals submitted by SiD and ILD~\cite{Aihara:2009ad,Abe:2010aa} were down-selected for validation. The two detector concept 
groups then presented their Baseline Designs in Volume 4 of the ILC TDR ~\cite{Behnke:2013lya}.

\subsection{ILC detector competence in Europe~\label{sec:det:competence}}
Europe has participated strongly in the global work on the ILC detector concepts and 
related technological developments. Initiated by the TESLA project at the start 
of the century, an international effort was put in place to develop the needed 
detector technologies. This is augmented by strong national support, but also by 
European programs like the EUDET initiative (2006-2010), the AIDA program 
(2011-2015) or the AIDA-2020 program (2015-2020). In Europe, both CERN and DESY 
act as a hub of activity and provide common infrastructures and facilities 
including test beams. Programs in Asia, in particular in Japan, and in the 
Americas have equally played an important role in the ILC concepts. 


 \begin{table}[htbp]

\begin{tabular}{l |c | c |c |c |c |c |c |c |c |c |c |c |c|}
           & \rot{ \bfseries Belgium} & \rot{ \bfseries CERN} & \rot{ \bfseries DESY} & \rot{ \bfseries Czech Republic} & \rot{ \bfseries France} & \rot{ \bfseries Germany} & \rot{ \bfseries Israel} & \rot{ \bfseries Netherlands} & \rot{ \bfseries Norway} & \rot{ \bfseries Poland} & \rot{ \bfseries Serbia} & \rot{ \bfseries Spain} & \rot{ \bfseries UK} \\[5pt]\hline 
\bfseries Vertexing  &   & \Checkmark 	 &\Checkmark	&\Checkmark		 &\Checkmark	  &\Checkmark	    & 	     & 		   & 	    &\Checkmark	     & 	     &\Checkmark	     &  \Checkmark \\[5pt]
\bfseries Tracking   & & \Checkmark	 &\Checkmark	& 		 &\Checkmark	  &\Checkmark	    & 	     &\Checkmark	 	   & 	    & 	     & 	     &\Checkmark      &  \Checkmark \\[5pt] 
\bfseries Calorimetry& \Checkmark & \Checkmark	 &\Checkmark	&\Checkmark		 &\Checkmark	  &\Checkmark	    &\Checkmark	     & 		   &\Checkmark	    &\Checkmark	     &\Checkmark	     &\Checkmark      &  \Checkmark \\[5pt] 
\bfseries MDI     &   & \Checkmark	 &\Checkmark	& 		 & 	  & 	    & 	     & 		   &\Checkmark	    & 	     & 	     & 	     &  \Checkmark \\ [5pt]
\bfseries Integration& &\Checkmark	 &\Checkmark	& 		 &\Checkmark	  & 	    & 	     & 		   & 	    & 	     & 	     &\Checkmark	     &    \\[5pt]\hline
\end{tabular}
\caption{An overview of recent activities in the area of ILC-related detector R\&D and integration in Europe based on a 2015 survey~\cite{FusterECFA2016}.}
\label{tab:PrePrep:detectors}
\end{table}

A summary of recent activities in ILC-related detector R\&D in Europe is given in Table~\ref{tab:PrePrep:detectors}~\cite{FusterECFA2016}. 
This table includes activities for CLIC, but not other more generic detector R\&D that can be applied to the 
ILC detectors. The strength in R\&D as well as the work on the detector concepts has put Europe in a strong 
position to contribute significantly to the future ILC detectors.

The technological developments have been described in detail in~\cite{ILCforESS}. 
The detectors at the ILC are very different from those at the LHC in that they 
can focus much more strongly on precision measurements. This is possible since 
the environment at an electron positron collider is much more benign than at a 
hadron collider. Radiation hardness hardly plays a role in designing the 
detectors. Multiple interactions, which are a major challenge at the LHC, are 
essentially not present. There is no trigger necessary and every crossing with collisions present will be recorded. Due to the long gap between the ILC bunch trains power pulsing of the electronics is possible leading to significant savings in the overall power and material budget.
These different boundary conditions allow a detector 
optimisation complementary to the one at the LHC. A focus on detailed event 
reconstruction, on high-precision vertexing and high-precision high-efficiency 
tracking and highly detailed imaging calorimetry is possible. As a guiding 
principle, Particle Flow has been widely accepted and is used by both SiD and ILD 
to characterise the performance of the complete systems. 

European groups have played strong roles in nearly all areas of detector design 
and development for the ILC. Particle flow calorimetry as a core element of ILC detectors,  
has been strongly pushed by European groups, mostly within the CALICE collaboration. Many developments on 
high-precision vertexing, in particular in the area of Monolithic Active Pixel 
Sensors (MAPS) technology, have come from European groups. The development of 
high-precision gaseous detectors based on micro-pattern gaseous detectors has 
been dominated by European groups. Last but not least, Europe has played a 
central role in the Machine-Detector-Interface and integration issues of 
both detectors.

\subsection{ILC Detector preparation phase activities in Europe~\label{sec:det:prepphase}}
The community is well prepared to move forward quickly once the ILC turns into a 
real project. We anticipate that detector construction will follow a path 
similar to that of the LHC detectors. Once the ILC laboratory has been formed, 
detector collaboration will formally start and develop concrete proposals based on the existing 
work for detectors at the ILC. To govern and organise this,  
strong central laboratory support will be essential. The host country would have 
to play a special role in this, but strong regional centres will also be very 
important. 

In Europe, CERN would be a natural candidate for such a regional 
centre, but national laboratories like DESY or others could also play this role 
for parts of the detector. Examples for such more regional centres within the 
LHC are the Detector Assembly Facility at DESY, used for the construction of 
parts of the upgrades to the LHC detectors, the CMS centre at Fermilab in the 
United States, or the CERN neutrino platform towards a European role in the long 
baseline experiments DUNE and HyperK. 

In contrast to the accelerator, within the overall guidelines defined by the host organisation the collaborations themselves will define and institute the needed structures to design and build the detectors. Strong support from the host 
organisation on all the issues of interfaces between the detectors and the accelerator is however essential. 

During the preparation phase, there are four major milestones for the
detectors.
\begin{itemize}

\item{\bfseries Optimisation} After the approval of the ILC the design of the 
existing detector concepts will need to be reviewed and refined. We expect that 
the community supporting these detectors will grow substantially. The design 
will also profit from the recent experiences of the LHC upgrade program, and 
other major detector construction projects. 

\item{\bfseries Integration into the ILC}
After the choice of the final location of the interaction region of the ILC, the 
detector designs need to be adapted to the
conditions of the site in terms of hall size, transport capabilities and
assembly space. Even though significant preparatory work on these questions is 
already happening, much more will be required. In particular the experience from 
Europe and from CERN will be invaluable.

\item{\bfseries Prototyping and Validation}
The ILC detectors have already reached an impressive level of maturity - more so 
probably than at any other large-scale HEP project at a similar stage in the 
past. Nevertheless the final designs and decisions will need a very vigorous and 
high-quality testing and prototyping program, to demonstrate the technical 
readiness, and to provide the basis for final technology decisions. This will 
require access to test beam and testing infrastructures in Europe and beyond, 
and will put particular demands on the test beam installations at CERN and at 
DESY. This process will continue well beyond the stage of the technical design 
report and approval of detectors, into the construction phase.

\item{\bfseries Technical Design Report for the detectors}
At the end of the preparatory phase, ILC detector concepts will present 
detailed technical design reports to the community and the funding bodies. Based 
on these reports the final decisions on which detectors will be build will be 
taken. 
\end{itemize}

Besides strong national support, the structuring of the European contribution should benefit from corresponding funding in upcoming calls by the European Union.

\subsection{\label{sec:det:constructionmodel} Estimation of a
European in-kind contribution to the ILC detectors}

In high-energy physics, the financial contribution to the detector
construction and operation is typically assumed to be proportional to the number of authors of
the detector collaboration. In the LHC experiments, Europe accounts for about 50\% of the members, 
in non-European experiments like Belle-II, European groups contribute about a 1/3 share. Most likely 
an experiment at the ILC would be closer to a Belle-II-like model, though, given the large interest 
in the community, possibly somewhere in between Belle-II and the LHC. 

Taking the cost numbers from the ILC TDR~\cite{Behnke:2013lya}, the European share for the detector construction 
is in the order of 270~M\euro{}, which is comparable to the cost for the ongoing upgrade of the LHC detectors for the HL-LHC. 
At the moment, some 57 institutions from 14 European countries have expressed 
an interest in participation in ILC-related detector work, a number that will increase after the ILC approval.

\subsection{\label{sec:det:Organisation} Organisation of the detector activities}
Traditionally, European groups have participated in projects outside CERN using bilateral 
agreements between the individual European countries and the host nation. One example is the European 
participation in the B-factories at SLAC and KEK or in the Tevatron programme at 
Fermilab.

More recently, the European participation in the long-baseline neutrino 
programme at Fermilab (LBNF/DUNE), while still being negotiated in the 
traditional bilateral way, has been augmented by the European Neutrino Platform 
hosted by CERN, which offers technical infrastructure and support for European 
groups working on detector and other contributions to long-baseline neutrino 
projects. CERN is also now formally a member of the DUNE collaboration.
In any case, as already outlined in Section~\ref{sec:acc:org}, our current assumption is that 
CERN will play a leading role in the European participation in the ILC, 
with strong support by the national laboratories.

\subsection{Computing}
As has been pointed out in the overall document on ILC~\cite{ILCforESS}, computing and data 
handling have been key to the successful feasibility study for the ILC. The ILC community 
has from the start invested in generic and widely usable tools. Starting from 
the initial development of a generic and detector-independent data model, the 
LCIO~\cite{bib:lcio} framework, a powerful and at the same time comparatively simple software 
infrastructure has been developed. All parts needed to simulate and to analyse 
events from the ILC are available to the community. A key motivation for investing
in this software and computing infrastructure has been to enable realistic and reliable 
studies on the science potential of the ILC. The Linear Collider Software is shared by ILC and 
CLIC, the concept groups, and most of the R\&D groups working on linear-collider-related 
studies. 

Europe has been a driving force in this endeavour. Strongly 
supported by both DESY and CERN and thanks to EU support in programs like EUDET, 
AIDA and AIDA-2020, the Linear Collider Software has now been expanded for use at test 
beam experiments and beyond the ILC     community.

In contrast to LHC computing, where a lot of the efforts are driven by pure 
size and data volume, the anticipated data rate from an ILC experiment is 
smaller by at least one order of magnitude. The computing architecture needed for 
ILC will follow the one for the LHC and while significant resources are needed 
to serve the experiments, these demands could already be met today without problems. 
The major computing challenge at the ILC will be the development and provision 
of advanced reconstruction and analysis tools, which will be capable of meeting 
the very challenging precision goals. In many cases, the requirements 
for these systems will go significantly beyond LHC/HL-LHC, 
and will require significant R\&D. Europe is in an excellent position to play a 
leading role in this part of the ILC project, based on its experience in LHC 
computing, and its contributions to ongoing linear-collider computing. 

\section{\label{sec:discussion}Discussion}
The ILC with a scientific program as outlined in~\cite{ILCforESS}, ranging from 
an initial stage at 250~GeV centre-of-mass energy, and reaching to higher 
energies up to around 1~TeV, will make major contributions to our understanding of the universe. 
It will be a large-scale international facility, which will 
complement and extend the science done at the LHC. It will pave the way - 
primarily through precision studies of the Standard Model - towards answering 
very profound questions on the nature of our universe. The ILC will be central to 
define the route particle physics should take after the LHC, and thus not only 
make major scientific contributions, but will help to shape the future of the 
world-wide HEP program. Strategy processes on the national and 
international level have repeatedly come to the conclusion that electron-positron collisions 
are at this time essential for progress in our field. 


We have described above the degree of preparation and of community involvement 
in the development of the accelerator and of the detectors at the ILC. This 
community is strongly committed - as shown by decades-long investment in the 
R\&D and prototyping of technologies - to build the most advanced and best 
possible accelerator and detector. The community is already very sizeable as can be seen from the list of signatories of the ILC Technical Design Report which includes 2400 signatories, 48 countries and 392 Institutes/Universities. In Europe, the community has united behind the 
series of  European-funded projects like (EUROTeV, CARE, ILC-HiGrade, E-JADE, EUDET, AIDA and AIDA-2020) and SCRF construction projects like the E-XFEL and ESS. This has 
resulted in Europe playing a very visible leadership role in the development of the ILC.



\subsection{\label{sec:discussionPol}Political synergy between Japan and Europe}
Interest in the ILC in Japan has been steadily growing for many years. The perspective of hosting the infrastructure in the country is promoted by political (Diet Federation, Tohoku district)
and industrial (AAA consortium) entities and finds strong support in the scientific community (KEK,
HEP community, JAHEP). The project is being examined in detail by bodies charged by the Japanese government. The scientific interest and political
engagement of partner countries is a major concern for the Japanese authorities. This
applies  to Europe in particular because of its expertise, capacities and involvement.

The political plurality of Europe necessitates an initial approach by the Japanese
authorities via bilateral discussions with individual countries, where ILC may appear in a broader
landscape embracing other advanced technology topics of mutual interest. This concept was first applied
in January 2018 when an official Japanese Delegation visited France and Germany. The Delegation
was composed of members belonging to the Parliament, to MEXT, to the AAA industry-academy
consortium, to the Tohoku district governance and to the HEP community. Meetings took place at
the French and German Parliaments and Ministries of Research. Some of the Delegation members took the opportunity
to visit CERN's infrastructures. The plan was next to extend such discussions to other countries as well as European
Institutions. This trip of the Japanese Delegation to France and Germany was complemented by several reciprocal
visits of French and German Parliament members in Japan during 2018, where the interest of
both countries in the ILC was explicitly expressed in the presence of a broad panel of Japanese political
representatives assuming a contribution of Europe to the total construction cost in the order of 20\% of the accelerator. This quantity matches the 1/3 contribution to the accelerator, excluding the civil engineering and infrastructure that are considered host responsibilities, and to the detectors as has been discussed in previous sections.



\subsection{\label{sec:discussionOrg} Organization of an European contribution}
Sections \ref{sec:acc:org} and \ref{sec:det:constructionmodel} give an initial understanding on how the future European contribution
to the ILC project could be managed for both the accelerator and the detectors. It is proposed that CERN, with strong support from the national laboratories (DESY, Saclay, etc.) will play a leading role in the European participation. 

In addition, bilateral agreements between Japan and interested European countries should also exist for specific studies, developments or deliverables. A similar scheme has already been put successfully in place for the European participation in the long-baseline neutrino program at Fermilab (LBNF/DUNE).
\vspace{0.5cm}
\subsection{\label{sec:discussionInd}Leveraging the expertise and the production capabilities of European industry}
With its strong academic and industrial involvement in the realisation of the E-XFEL, Europe is particularly well prepared to contribute to the ILC construction. Based on its comprehensive expertise and its demonstrated production capability for all main components of such a machine, European industry is poised to play a central role in the project realisation. 

The example of the E-XFEL illustrates how technological innovation driven by particle physics impacts research in other domains, such as biology, pharmaceutics and material science. Moreover, the technological advances provided by the R\&D for the ILC and E-XFEL are now being used to realise other infrastructures such as the ESS.
The achieved high performances of the E-XFEL technology resulted also in its adoption in the USA (LCLS-II at SLAC) and China (SHINE in Shanghai) for high-brilliance light sources, to be constructed in part by European companies which have realised the E-XFEL. These companies are well aware of the ILC project and follow its evolution with strong interest. Interactions with industry take place regularly e.g. at the international linear collider conferences in dedicated industrial sessions. These regular sources of exchange emphasise that the attractiveness of the ILC for European industry is multi-faceted; 
it is seen as a way to maintain technological leadership in this area. 

\subsection{Conclusion}
Europe has demonstrated leadership in both the accelerator and the detector part of the ILC project. Scientifically the case has been shown to be stronger than ever. European industry is technologically well qualified and prepared to construct important parts of the ILC accelerator and detectors. The ILC project is comparable in size to the LHC. The European involvement in a Japanese-hosted 250~GeV ILC, as discussed in this document, is around 20\%. The ILC project fits well with the HL-LHC time-line moving into major construction around 2025 and will play a crucial role in encouraging a new generation of researchers to enter particle physics while maintaining present expertise. A strong European linear collider community exists which is eager to participate in the project and to strongly contribute to its realisation.

\bibliography{ILC-ESU-refs}

\onecolumngrid
\newpage

\appendix

\addcontentsline{toc}{part}{Appendix}

\section*{\label{Appendix3} \Large{Appendix A: List of supporting documents} }
\begin{itemize}
\item
ILC TDR documents;
\item
ILC general overview, being specifically produced for the European Strategy Process;
\item
European ILC Preparation Plan (EIPP), produced under the E-JADE project;
\item
Linear collider Detectors R\&D Liasion Report;
\item
Green ILC project: reports and web page;
\item
Letter from the KEK’s ILC Planning Office.

\end{itemize}

\textbf{Supporting documents web page:} 


\url{https://ilchome.web.cern.ch/content/ilc-european-strategy-document}

\section*{\label{Appendix4} \Large{Appendix B: Glossary} }
\begin{itemize}
\item
\textbf{AAA:} The Japanese Advanced Accelerator Association promoting science and technology (\url{http://aaa-sentan.org/en/association/index.html}).
\item
\textbf{AIDA:} Advanced European Infrastructures for Detectors at Accelerators. AIDA was funded by the EU under FP7 (\url{https://aida-old.web.cern.ch/aida-old/index.html}).
\item
\textbf{AIDA-2020:} Advanced European Infrastructures for Detectors at Accelerators. The successor of AIDA; AIDA-2020 is funded by the EU under Horizon2020 (\url{http://aida2020.web.cern.ch/}).
\item
\textbf{CALICE Collaboration:} R\&D group of more than 280 physicists and engineers from around the world, working together to develop a high granularity calorimeter system optimised for the particle flow measurement of multi-jet final states at the ILC running, with centre-of-mass energy between 90 GeV and ~1 TeV (\url{https://twiki.cern.ch/twiki/bin/view/CALICE/WebHome}).
\item
\textbf{CARE:} Coordinated Accelerator Research in Europe. CARE was funded by the EU under the FP6 programme.
\item
\textbf{E-JADE:} The Europe-Japan Accelerator Development Exchange Programme. E-JADE is a Marie Sklodowska-Curie Research and Innovation Staff Exchange (RISE) action, funded by the EU under Horizon2020 (\url{https://www.e-jade.eu/}).
\item
\textbf{EUDET:} Detector R\&D towards the International Linear Collider. EUDET was funded by the EU under the FP6 programme (\url{https://www.eudet.org/}).
\item
\textbf{European XFEL:} The European X-Ray Free-Electron Laser Facility (European XFEL) at DESY (Hamburg, Germany) (\url{https://www.xfel.eu}).
\item
\textbf{EUROTeV:} European Design Study Towards a Global TeV Linear Collider. EUROTeV was funded by the EU under the FP6 programme (\url{https://www.eurotev.org/}).
\item
\textbf{ICFA:} International Committee for Future Accelerators (http://icfa.fnal.gov/).
\item
\textbf{ILC-HiGrade:} International Linear Collider and High Gradient Superconducting RF-Cavities. ILC-HiGrade was funded by the EU under the FP7 programme (\url{https://www.ilc-higrade.eu/}).
\item
\textbf{JAHEP:} Japanese Association of High Energy Physics.
\item
\textbf{Japanese National DIET:} The National Diet is Japan's bicameral legislature. It is composed of a lower house called the House of Representatives, and an upper house, called the House of Councillors.
\item
\textbf{LCLS-II:}  The hard X-ray free-electron laser at SLAC (Stanford, USA)(\url{https://portal.slac.stanford.edu/sites/lcls-public/lcls-ii/Pages/default.aspx}).
\item
\textbf{MEXT:} Ministry of Education, Culture, Sports, Science and Technology (\url{http://www.mext.go.jp/en/}).
\item
\textbf{SHINE:} Hard X-Ray free electron laser facility in Shanghai.

\end{itemize}

\end{document}